# A graph-based multimodal framework to predict gentrification

Javad Eshtiyagh[*]
Baotong Zhang[*]
Yujing Sun[*]
Linhui Wu[*]
Zhao Wang[*]

October 24th, 2023

**Abstract**

Gentrification—the transformation of a low-income urban area caused by the influx of affluent residents—has many revitalizing benefits. However, it also poses extremely concerning challenges to low-income residents. To help policymakers take targeted and early action in protecting low-income residents, researchers have recently proposed several machine learning models to predict gentrification using socioeconomic and image features. Building upon previous studies, we propose a novel graph-based multimodal deep learning framework to predict gentrification based on urban networks of tracts and essential facilities (e.g., schools, hospitals, and subway stations). We train and test the proposed framework using data from Chicago, New York City, and Los Angeles. The model successfully predicts census-tract level gentrification with 0.9 precision on average. Moreover, the framework discovers a previously unexamined strong relationship between schools and gentrification, which provides a basis for further exploration of social factors affecting gentrification.

[*] University of Chicago, Division of Social Sciences (Computational Social Science)

**1. Introduction**

Gentrification in almost every US major city has become a trending media topic not only in the likes of The New York Times and Forbes but also in academic literature, indicating the significance of the phenomenon. Turning into a hot topic has resulted in various definitions and interpretations of gentrification. The term gentrification was first coined by Ruth Glass in 1964 while she was describing the neighborhood change in London. She defined gentrification as a process causing "all or most of the working class occupiers [to be] displaced and the whole social character of the district [to be] changed" (Glass 1964).

Gentrification enhances the socioeconomic development of the affected areas and is typically accompanied by building renovations and amenity enhancements (Zukin 1987). However, gentrification could also result in severe challenges for low-income residents. For example, due to the increasing demand for housing in a gentrifying neighborhood, low-income tenants either have to spend more on housing or move out of the area (Vigdor, Massey, and Rivlin 2002; Atkinson 2002). Policymakers are concerned with balancing between the benefits of urban revitalization and the displacement of low-income families (Levy, Comey, and Padilla 2006). The efficacy of gentrification-related policies is a function of how early they are administered. Hence, predicting which parts of the city are susceptible to gentrification would enable policymakers to develop effective customized policies that can reduce potential side effects prior to or at the early stages of gentrification.

In addition to the well-known socioeconomic and geographic factors (Kolko 2007), researchers have also discovered variables such as transportation accessibility (Padeiro, Louro, and da Costa 2019; Baker and Lee 2019), public investment (Zuk et al. 2018), building characteristics (Helms 2003), and urban natural resources (Rigolon and Németh 2020) to be effective in identifying and predicting gentrification. However, the majority of the quantitative studies that predict gentrification using these variables rely on long-term economic, social, and demographic panel data, which are typically collected and updated over five-year or ten-year periods.

In recent years, computational social scientists have been developing machine learning models to predict gentrification. Reades, De Souza, and Hubbard (2019) apply random forests with economic and demographic features to predict gentrification in London. Naik et al. (2017) explore a large dataset of Google Street View images to discover neighborhood change over time. Ilic, Sawada, and Zarzelli (2019) conduct a detailed image analysis to predict whether individual buildings will be renovated or replaced by new ones. These models significantly outperform traditional statistical studies, but given the complexity of gentrification, they are still limited by available data sources and input features, i.e. dependence on temporal and low-resolution data.

As illustrated in Figure 1, we propose a graph-based multimodal deep-learning framework to predict gentrification at the census tract level. Specifically, we extract features from multiple data sources: building footprints from satellite images, socioeconomic features from the US census database, and network features by constructing the urban networks from essential facilities (e.g.

schools, hospitals, and subway stations). With the multiple sources of features, we carry out two types of parallel experiments: (a) take all the available features to construct a unified urban network and then apply GraphSAGE to predict gentrification; (b) feed each source of features into a single model (e.g. Random Forest with socioeconomic features, Graph Convolutional Network (GCN) for the urban network) and explore their predictive power separately.

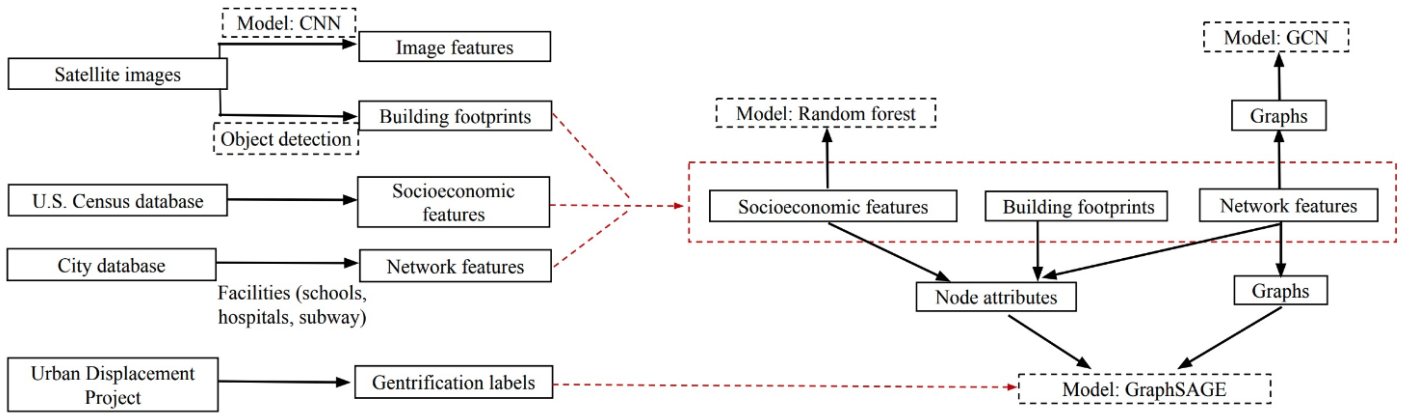

Figure 1: The graph-based multimodal framework

We conduct experiments with the three most populous cities in the United States: Chicago, New York, and Los Angeles. Results indicate that: (a) with diverse sources of features serving as node attributes in the constructed urban network, the proposed graph-based framework achieves high predictive ability with an average of 0.9 AUC; (b) among the single models trained with each individual source of features, the urban network built from tracts and nearby schools shows the most impressive prediction performance, suggesting the strong relationship between schools and gentrification.

To summarize, this work has two main contributions: First, we believe there is a gap in the literature in examining the relationship between gentrification and facilities such as schools, hospitals, and subway stations. In this study, we propose a new way of constructing urban networks from tracts and facilities, as well as incorporating socioeconomic variables and satellite image features into the network. This new method uncovers a remarkable relationship between schools and gentrification that calls for further investigation.

Second, prior works tend to predict gentrification by detecting temporal patterns from longitudinal data. However, such data are not always accessible and the patterns typically emerge after gentrification has already begun, which is a late signal for policymakers. Although gentrification is a dynamic process across a couple of years, some gentrification predictors could be stable across these years. For example, if a census tract has facilities like schools, hospitals, and public transportation, then this tract, if not already gentrified, will very likely to be gentrified in the near future. The existence of such facilities is generally stable across time and can be used as predictors of future gentrification. Based on such rationale, we propose a practical model that takes data from only one historical year to make early predictions of gentrification at the census tract level.

## 2. Literature review

Traditional work in identifying and predicting gentrification includes both qualitative and quantitative methods. Some researchers take the qualitative approach to investigate social and cultural patterns that occur due to changes in racial composition, income distribution, or type of business (e.g. local stores vs chain stores) (Barton 2016; Curran 2007; Zukin et al. 2009). The majority of researchers take the quantitative approach to analyze demographic and economic datasets. Such studies typically rely on identifying temporal patterns from longitudinal data (Levy 1986). However, such analysis has two issues. First, they rely on patterns that emerge only after gentrification has begun in an area. Second, due to limitations in obtaining the required microdata, analyses based on these methods result in low accuracy.

To alleviate the existing problems, some social scientists propose to make predictions based on certain neighborhood characteristics. For example, Padeiro, Louro, and da Costa (2019) and Baker and Lee (2019) found that the availability of transportation services is predictive of gentrification. Helms (2003) discovered that the age and architectural style of buildings in a neighborhood have significant correlations with gentrification. Other amenities, such as access to parks, have also been discovered to be predictors of gentrification (Rigolon and Németh 2020). These studies take a step in the direction of solving the above issues but usually end up with insufficiently accurate predictions.

In recent years, scholars have turned towards exploring machine learning methods to predict gentrification. One of the earliest models was implemented by Reades, De Souza, and Hubbard (2019) that applied a random forest model with socioeconomic features to predict gentrification in London with data spans over a ten-year period. Liu et al. (2019) showed that unsupervised learning can also be effective in making predictions using socioeconomic data at the census tract level. Recently, the IBM Data Science Elite (Copty 2021) team announced a new collaboration with the Urban Institute with the aim of predicting gentrification years before it occurs.

In the emerging line of predicting gentrification, some researchers are working on analyzing image data to predict neighborhood change. Naik et al. (2017) and Ilic, Sawada, and Zarzelli (2019) use Google Street View images to make predictions. These models have promising results, however, they are currently limited to predicting whether individual buildings become renovated or rebuilt.

Despite the improvements over traditional methods, there is still space for further improvement. Many of the well-performing models continue to rely on temporal patterns observed in gentrification. Such models are not applicable to make predictions before gentrification begins. Some models also use high-frequency and high-resolution panel data that are not usually available in many contexts. Moreover, the accuracy of available models is not high enough to be confidently accepted by policymakers. Lastly, beyond the commonly explored factors, there are other aspects of urban life connected to gentrification that are not taken into account in the existing models. Motivated by these observable gaps, we propose a graph-based multi-modal framework to effectively predict gentrification using stable features (e.g. facilities such as schools, hospitals, and subway stations) from one year of data.

## 3. Dataset

This section introduces the data that we use for training and testing the proposed framework. We collect data from the three most populous cities in the United States: Chicago, New York, and Los Angeles. For each city, we collect satellite images, socioeconomic data, locations of facilities (i.e., schools, hospitals, and subway stations), and gentrification labels at the census tract level.

*Satellite images:* We download satellite images from Google Earth Engine[1] for each tract. These raw images are then cropped and padded to the same size and scale. Figure 2 shows an example of the satellite image for a tract in Chicago.

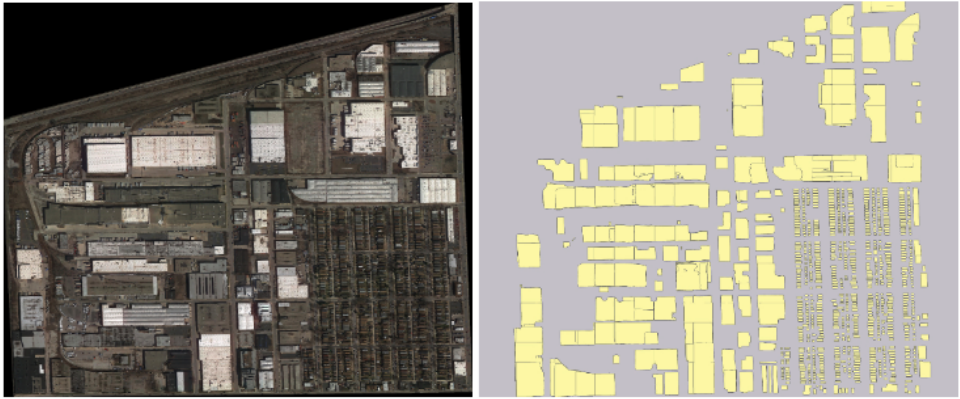

Figure 2: An example of the satellite image and object detection result for this image

*Socioeconomic data:* We collect data from the US Census database (United States Census Bureau 2022) and select features that capture the social, economic, and demographic dimensions of each tract. Details of the feature selection process are explained in §4.2.

*Essential facilities:* Previous works have shown that access to essential facilities (e.g., parks and transportation services) is related to gentrification (Rigolon and Németh 2020; Baker and Lee 2019). In this study, we continue to explore how accessibility to essential facilities is related to gentrification. We experimented with accessibility to various facilities such as police stations, fire stations, parks, schools, hospitals, subway stations, etc. Among all the tested ones, schools, hospitals, and subway stations showed the strongest correlations with gentrification.

---

[1] https://earthengine.google.com/

Table 1: Number of essential facilities and tracts collected for each city.

|  | Chicago | New York | Los Angeles |
|---|---|---|---|
| # gentrified tracts | 155 | 489 | 272 |
| # non-gentrified tracts | 627 | 1,459 | 1,380 |
| # schools | 745 | 1,817 | 1,213 |
| # hospitals | 48 | 78 | 95 |
| # subway stations | 137 | 473 | 164 |

*Gentrification labels:* The labels are collected from the Urban Displacement Project (Chapple, Thomas, and Zuk 2021), an organization that conducts research to understand the nature of gentrification, displacement, and exclusion. It determines the gentrification label of each tract based on economic and demographic changes (e.g. changes in average household income and housing cost). This project lists 10 categories describing the level of gentrification and we group them into two classes for exploration: (a) gentrified/gentrifying, which includes areas at risk of imminent gentrification, early/ongoing gentrification, and advanced gentrification; and (b) non-gentrified, which includes all the remaining areas. It is possible to group these categories into multiple classes and explore more fine-grained levels of gentrification in future work. In this current work, we are mainly focused on exploring features that could distinguish between gentrified and non-gentrified areas and thus serve as predictors of gentrification.

The most recent gentrification labels we can retrieve from the Urban Displacement Project are from 2017 for Chicago, 2016 for New York, and 2018 for Los Angeles. Thus, we collect all the discussed data for each city in the year when the gentrification label is available. We then construct models to predict the gentrification status of each tract in the same or near future year. As discussed before, our framework is designed to utilize features that are stable across years. For example, the essential facilities (e.g., schools, hospitals, subways) in a neighborhood do not significantly change over a 5 or 10-year period. Thus, these features are consistent as long as they are extracted from recent years. As a result, we can use these stable features to predict gentrification for tracts that have unknown gentrification status in the current or future years. Table 1 shows a summary of the collected data for each city.

## 4. Framework and input features

In this section, we first introduce the features that we obtain from the raw data and explain their relationship with gentrification. We then propose a novel method to construct urban networks from census tracts and urban facilities. Lastly, we present our graph-based multimodal framework that utilizes various modes of features to predict the likelihood of gentrification for each census tract.

*4.1 Satellite image features*

We are interested in exploring whether satellite images contain important information that could predict future gentrification. The recent advances in deep learning facilitate the process of learning complex details from images. Specifically, we do object detection with the pre-trained *Building Footprint Extraction* model (Esri 2021), which uses the *MaskRCNN* architecture (He et al. 2017) based on Arc API[2] to gather details from buildings in satellite images. For each tract, we extracted features such as the number, the circumferences, and the areas of the buildings.

*4.2 Socioeconomic features*

Previous studies have found that temporal patterns of socioeconomic changes, such as income, education, and rental prices, are predictive of gentrification (Reades, De Souza, and Hubbard 2019; Hwang and Sampson 2014). In this work, we are interested in the predictive power of socioeconomic features from a single year. First, raw socioeconomic data are collected from the Census Bureau's database. We then conduct feature selection with L1 regularization using a logistic regression classifier based on gentrification labels to select features that are correlated with gentrification status. Lastly, we run a correlation analysis among the selected features to remove those that are multicollinear with each other. This process gives us 18 effective socioeconomic features for each city. Appendix 1 shows these features ranked by their coefficient magnitudes.

```
Algorithm 1: Constructing the Urban Network
Input: V = {T, S, H, C}, the set of nodes
Output: G = (V, E), the constructed urban network

 1: Let G = networkx.Graph()
 2: for t ∈ T do
 3:     G.add_node(t)
 4:     for P ∈ {S, H, C} do
 5:         distance = [ ]
 6:         for p ∈ P do
 7:             distance.append(distance between t and p)
 8:         end for
 9:         for p in sorted(distance)[: 2] do
10:             G.add_node(p)
11:             G.add_edge(t, p)
12:         end for
13:     end for
14: end for
15: return G
```

---

[2] https://doc.arcgis.com/en/pretrained-models/latest/imagery/introduction-to-building-footprint-extraction-usa.htm

*4.3 Constructing urban networks*

A major contribution of this study is to test whether access to schools, hospitals, and subway stations plays a significant role in determining the likelihood of an urban area being gentrified in the future. To the best of our knowledge, we are not aware of any prior work that considers urban networks to study gentrification. Here, we propose a novel method to construct urban networks from tracts and essential facilities.

We first collect the basic information (e.g. coordinates, name, and address) about the essential facilities from the city database of Chicago[3], New York[4], and Los Angeles[5]. We then construct urban networks based on the connections between tracts and facilities. Specifically, we construct the urban network by representing tract centroids and facilities as nodes and then add undirected edges between each tract and the two closest facilities. The weight of each edge corresponds to the direct distance between the center of the tract and the connected facility. We have experimented with multiple configurations of connecting tracts to the closest n facilities (n ranges from one to ten), and results show that connecting with the two closest facilities provides the optimal performance for predicting gentrification. We believe this is because connecting with two facilities provides the optimal level of detail about the urban networks.

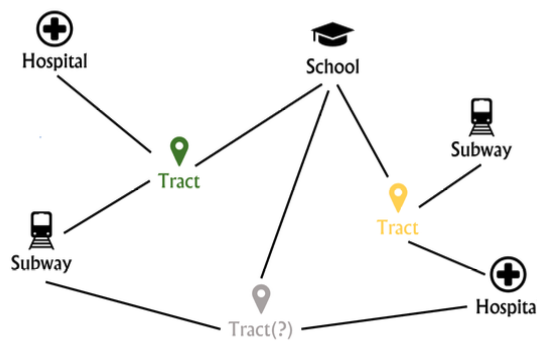

Figure 3: A sample graph of the constructed urban network. *Note: this is a simplified graph, each tract is connected with one facility (i.e., school, hospital, subway). The tract in green color is gentrified, the tract in yellow is non-gentrified, and the tract in grey is to be predicted.*

Formally, we describe the process of constructing the urban network in Algorithm 1 with the following notation: *G=(V, E)*, where *G* is a graph constructed by connecting nodes *V* with edges *E*; and *V={T, S, H, C}* is the set of four types of nodes: *T* for tracts, *S* for schools, *H* for hospitals, and *C* for subway stations. Figure 3 shows a sample subgraph of the constructed network and Figure 4 illustrates the graph that connects tracts and hospitals in Chicago. Based on the data presented in table 1, in this case, the total number of nodes is #gentrified + #non-gentrified +

---

[3] https://data.cityofchicago.org/
[4] https://opendata.cityofnewyork.us/
[5] https://geohub.lacity.org/

#hospitals = 155 + 627 + 48 = 830 and the number of edges is equal to (155 + 627) × 2 = 1564, i.e. each tract is connected with two closest hospitals.

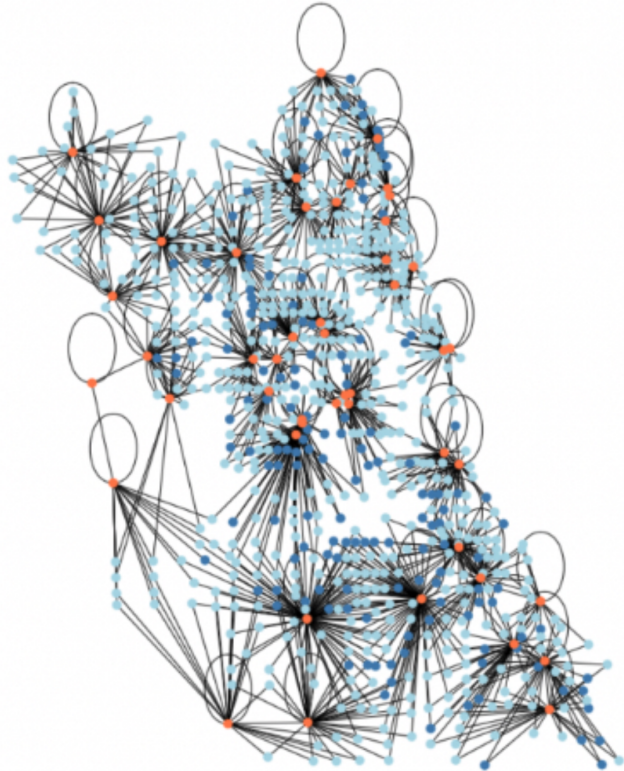

Figure 4: The urban network constructed with tracts and hospitals in Chicago. Each tract is connected to the two closest hospitals. Gentrified tracts are represented by dark blue, non-gentrified tracts by light blue, and hospitals by orange.

*4.4 The graph-based multimodal framework*

In this study, we try to combine multiple sources of features in the multi-modal framework. We first create the urban networks following Algorithm 1 and then add features as node attributes and assign gentrification labels to each node in the urban network. Finally, we feed this urban network into *GraphSAGE*, a framework for inductive representation learning on large graphs (Hamilton, Ying, and Leskovec 2017), and train a binary classifier to predict the gentrification label for each node. Specifically, the features that serve as node attributes are as follows:

- Image features of buildings obtained from object detection: the number of buildings, the shape lengths, and the footprint of each building.
- Social, economic, and demographic features obtained from census data.
- One-hot encoded facility type (i.e. school, hospital, and subway station).
- The shortest paths, i.e. the number of nodes between each tract and its closest facilities, provide another measure for how accessible facilities are to each tract. To get the shortest path, we form a graph by connecting each node to its neighboring nodes and then applying Dijkstra's algorithm using the *NetworkX* package (Hagberg, Swart, and S Chult 2008). Figure 5 illustrates this in the case of Chicago. Adding shortest paths helps the model to focus not only on the direct distance between tracts and facilities provided by the graph shown in figure 4, but also on the relative position of nodes with respect to each other. In other words, the model also takes into account which tracts a resident has to travel through to get to the closest facility.

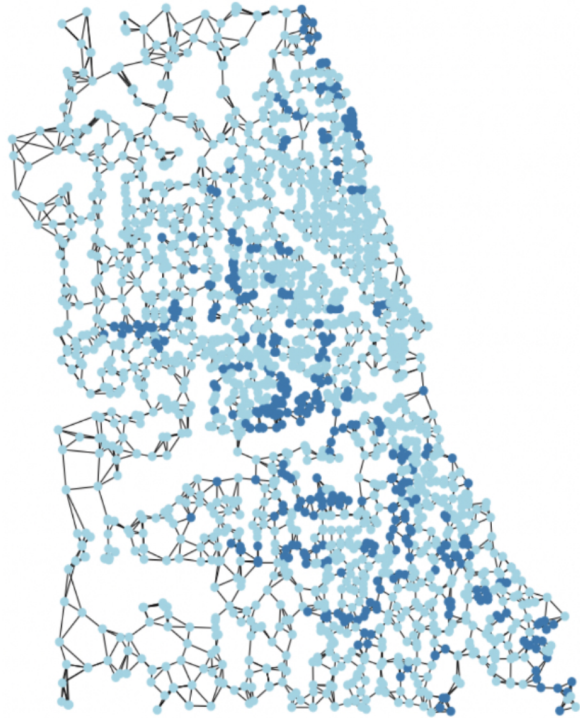

Figure 5: The network constructed for calculating the shortest paths in Chicago.

*4.5 Other models*

In addition to the multimodal framework, we experiment with a random forest model for only socioeconomic data and a graph neural network (GCN) for only the urban graphs besides applying GraphSAGE to subsets of input features. The use of random forest and GCN is motivated by the fact that these models outperform GraphSAGE when analyzing tabular or graph features separately on their own. The results and the comparison to GraphSAGE are presented in section 5.

## 5. Experiments and results

We first conduct experiments to evaluate the performance of the graph-based multi-modal framework (i.e., GraphSAGE) with all features combined as well as only subsets of features.

Then, to better understand the predictive ability of each type of feature, we design individual random forest and GCN models to predict gentrification using each mode of data separately. We find that although GraphSAGE is optimal when both the graph and node attributes are present, random forest performs better when focusing only on socioeconomic features and GCN performs better when making predictions only based on urban networks. We run experiments in the three cities of Chicago, New York, and Los Angeles.

*5.1 Experimental settings*

In the experiments, each tract serves as one data point. We randomly split the dataset and use 80% of the data points for model training and 20% for testing. We do n-fold cross-validation[6] with training data to find the best hyper-parameter settings for each model and then test model performance on the testing set. We repeat each experiment 10 times and report the average precision, recall, F1, and AUC scores for every model[7]. These scores provide a thorough understanding of the predictive ability of each machine learning model from multiple perspectives (e.g., accuracy, completeness). In the following contents, we use precision as an example to describe the model performance, and the comparative performance is consistent with other measurements such as recall, F1, and AUC scores.

*5.2 GraphSAGE for the urban networks constructed with various features*

Following the instructions in §4.3, we construct one urban network for each city and feed each urban network into a *GraphSAGE* model. The constructed GraphSAGE model has three layers with dimensions 16, 4, and 2. We ensure the loss curves converge after training 400 epochs with a batch size of 150.

Table 2: GraphSAGE model performance with different features

|  | Chicago | | | | New York | | | | Los Angeles | | | |
| --- | --- | --- | --- | --- | --- | --- | --- | --- | --- | --- | --- | --- |
| **Features** | **Precision** | **Recall** | **F1** | **AUC** | **Precision** | **Recall** | **F1** | **AUC** | **Precision** | **Recall** | **F1** | **AUC** |
| Tracts and facilities | 0.79 | 0.81 | 0.80 | 0.77 | 0.76 | 0.77 | 0.77 | 0.77 | 0.79 | 0.82 | 0.80 | 0.80 |
| Building footprints | 0.82 | 0.78 | 0.80 | 0.81 | 0.79 | 0.79 | 0.79 | 0.82 | 0.80 | 0.82 | 0.81 | 0.82 |
| Shortest path | 0.81 | 0.83 | 0.82 | 0.82 | 0.76 | 0.77 | 0.76 | 0.80 | 0.80 | 0.81 | 0.80 | 0.80 |
| Socioeconomic | 0.91 | 0.92 | 0.92 | 0.89 | 0.88 | 0.87 | 0.87 | 0.90 | 0.88 | 0.87 | 0.88 | 0.87 |
| All features combined | **0.92** | 0.91 | 0.91 | 0.92 | **0.89** | 0.89 | 0.89 | 0.90 | **0.87** | 0.88 | 0.88 | 0.87 |

We run the model with different feature sources serving as node attributes. Table 2 summarizes the results. As observed, the implemented graph-based model combining all sources of features has a remarkable performance in predicting gentrification. Specifically, it achieves precision scores of 0.92 for Chicago, 0.89 for New York City, and 0.87 for Los Angeles. Figure 6 further illustrates the ROC curves (Fawcett T 2006). These ROC curves show the classification performance for the three cities based on the true positive rate versus the false positive rate at various threshold settings. This evaluation method is used when the classification training data is not balanced. We observe similar trends in the recall, F1, and AUC scores (Area Under the

---

[6] https://scikit-learn.org/stable/modules/cross_validation.html
[7] https://scikit-learn.org/stable/modules/model_evaluation.html

[ROC] Curve). Comparing the model performance with only one source of attributes, socioeconomic features show better predictive ability than the shortest paths and building footprints. This is not surprising due to the large number of socioeconomic features fed into the model.

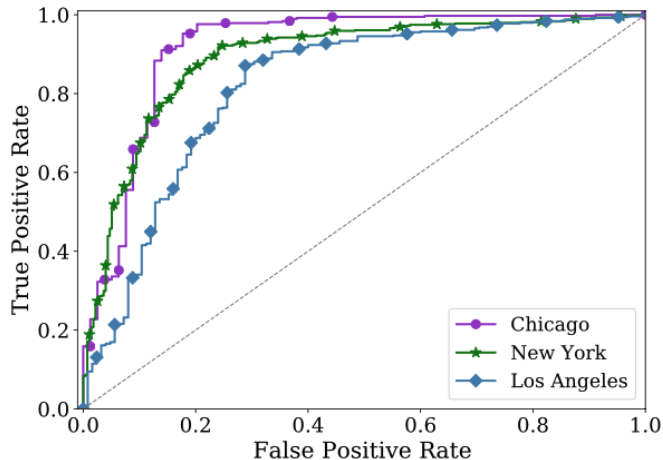

Figure 6: ROC curves for the GraphSAGE model performance with all features serving as node attributes in the urban network.

*5.3 GCN for urban networks constructed with each type of facility*

To better understand the predictive ability of each type of essential facility, we construct three separate networks for schools, hospitals, and subway stations for each city. The process for creating the urban network is similar to the one described in §4.3. Take the school network as an example, we create the urban network by representing tracts and schools as nodes and connecting each tract with the two closest schools. To analyze graphs without any node attributes, graph convolutional networks (GCNs) (Fey and Lenssen 2019) outperforms GraphSAGE. This is due to the fact that since GraphSAGE is meant to simultaneously learn from node attributes, it analyzes simplified embeddings of inputted graphs. Hence, we apply GCN models to predict the gentrification label for each node purely based on the urban network structure for each type of facility.

Table 3: GCN model performance with different urban networks

|  | Chicago | | | | New York | | | | Los Angeles | | | |
|---|---|---|---|---|---|---|---|---|---|---|---|---|
| Urban networks | Precision | Recall | F1 | AUC | Precision | Recall | F1 | AUC | Precision | Recall | F1 | AUC |
| School network | **0.87** | 0.88 | 0.88 | 0.81 | **0.85** | 0.85 | 0.85 | 0.83 | **0.88** | 0.89 | 0.88 | 0.77 |
| Hospital network | 0.73 | 0.77 | 0.75 | 0.55 | 0.70 | 0.75 | 0.70 | 0.56 | 0.79 | 0.83 | 0.80 | 0.58 |
| Subway network | 0.77 | 0.79 | 0.78 | 0.60 | 0.73 | 0.75 | 0.73 | 0.62 | 0.79 | 0.83 | 0.78 | 0.54 |
| Network with all facilities | 0.87 | 0.86 | 0.86 | 0.79 | 0.84 | 0.84 | 0.84 | 0.81 | 0.85 | 0.86 | 0.85 | 0.76 |

Table 3 presents the results for different cities and facility types. Hospitals and subway stations are predictive of gentrification, but most importantly, there appears to be a strong relationship between schools and gentrification. The GCN model trained on the school network achieves precision scores of 0.87, 0.85 and 0.88 in Chicago, New York, and Los Angeles respectively. One could form both technical and social hypotheses for explaining this discovered high predictive ability of schools.

One may argue technical reasoning for this observation as the model can learn valuable information from the neighboring nodes. In the constructed urban networks, visualized in Figure 3, relatively independent subgraphs are formed around each facility. Suppose we are interested in predicting the label of a tract in one of these subgraphs. The technical reasoning implies that the model would predict the tract to have the same label as the majority of the other tracts in that subgraph. With this reasoning, one can argue that since the number of schools in a city is larger than the number of hospitals and subway stations, the subgraphs formed around schools are smaller and the predictions are more accurate.

However, this technical reasoning cannot fully explain the observed predictive power. For example, we find that private schools achieve higher predictive power than public schools. Given only the technical reasoning, one would have expected the opposite because public schools have a significantly larger number. Furthermore, because the number of subway stations is larger than the number of hospitals, one would expect the subway GCNs to be much more effective than the hospital GCNs, but this is not the case in the results: the subway model results are only slightly more accurate than the ones from the hospital model in two of the cities and even less accurate in the case of Los Angeles.

The social explanation of the high prediction power of schools appears to be more plausible. Many households choose the location of their future homes based on education resources (Lareau and Kimberly 2014), thus a good school can itself contribute to the gentrification of a tract. Recall from the socioeconomic model that the average age in a tract and the number of households with school-age children are notable predictors of gentrification. This observation corroborates the hypothesis regarding the potential contribution of schools to gentrification and is supported by previous findings that most of the gentrifying households are families with young children (Moos 2016).

We believe that this discovery forms a basis for future study of the strong relationship between schools and gentrification.

*5.4 Model performance with other features*

Although the focus of this paper is to explore the relationship between essential facilities and gentrification, we also explored features suggested in prior works: socioeconomic features and satellite image features.

*Random forest classifier with socioeconomic features:* Having explored multiple traditional and deep learning models, we discovered that the random forest classifier with 10 estimators and "entropy" as node splitting criteria shows notable performance (Table 4). The model achieves precision scores of 0.81, 0.82, and 0.81 in the three cities suggesting that socioeconomic data contain important information that is predictive of gentrification, which is also supported by our findings in Table 2.

Table 4: Random forest classifier performance with socioeconomic data.

|  | Precision | Recall | F1 | AUC |
|---|---|---|---|---|
| **Chicago** | 0.81 | 0.84 | 0.81 | 0.83 |
| **New York** | 0.82 | 0.83 | 0.82 | 0.86 |
| **Los Angeles** | 0.81 | 0.84 | 0.80 | 0.81 |

For a deeper understanding of the socioeconomic features, consider the example of Chicago. Some of the predictive features we discovered are: (1) racial composition, which is aligned with previous studies investigating the dynamic relationship between racial composition and gentrification (Anderson and Sternberg 2013; Goetz 2010); (2) higher education resources, which is attractive to families seeking education resources (Desena and Ansalone 2009); (3) rent gap between neighborhoods, which means if a tract's rental rates are much lower than the rates in neighboring tracts then that tract is more likely to be gentrified in the future. Additionally, variables such as the number of households with children, healthcare, art, and entertainment services.

*CNN models with satellite image Features:* Researchers state that housing and neighborhoods are important indicators of gentrification (Helms 2003). We believe satellite images can provide great details about such indicators. So, we explored multiple CNN models, both tried to train our own model and fine-tune pre-trained models such as ResNet50. Our goal was to learn geographic community features that are expected to distinguish between gentrified and non-gentrified tracts, such as the street structure (e.g. the transportation network and road width), housing features (e.g., appearance), green space, and other valuable details. Unfortunately, we do not get convincing results from the CNN models. We conclude the reasons as follows: (a) limited training data: satellite images contain rich information about the tracts and we need a deep neural network to learn complex features from the image. However, the complex models are not well trained with limited data; (b) lack of temporal information: gentrification is a process of movement, and changes observed from the periodical data can provide great indicators. Although we successfully applied object detection results to accomplish our prediction task, more detailed image features remain unexplored (e.g., temporal change of building style and park design, or the age of the architecture). We believe the model performance could be improved by collecting more satellite images from multiple years and conducting a time-series analysis to learn the changing pattern of the geographic features.

*5.5 Summary*

In the above experiments, we evaluated GraphSAGE model performance with self-constructed urban networks added with traditional features (i.e., *socioeconomic and satellite image features*), GCN model performance with self-constructed facility-specific urban networks, and other model performance with each type of traditional features. Results show that our proposed graph-based multimodal framework implemented with GraphSAGE, achieves the best precision scores of 0.92, 0.89, 0.87 for Chicago, New York, and Los Angeles. GCN models with urban networks constructed from schools (precision scores of 0.87, 0.85, 0.88 for three cities) outperform hospital networks and subway networks for the three cities. Both GraphSAGE and GCN achieve higher precision than models using traditional features, suggesting that our self-constructed urban networks based on tracts and facilities serve as important gentrification explainers.

## 6. Conclusion and future work

To help local policymakers take early and targeted action in protecting underprivileged residents against gentrification, computational social scientists have been developing machine learning models to predict gentrification before it occurs at the neighborhood level. Building upon previous studies that relied on training models with socioeconomic data and image features, we propose a novel graph-based multimodal framework that predicts gentrification based on the constructed urban networks. More specifically, we construct urban networks based on the locations of census tracts and the two closest schools, hospitals, and subway stations to that tract. These urban networks are then combined with previously analyzed socioeconomic and satellite image features to achieve more accurate predictions.

In addition to discovering the predictive power of such networks, we were also able to compare the predictive abilities of different sources of features. This comparison further highlighted the connection between the self-constructed urban networks and the gentrification phenomenon.

In particular, the analysis sheds light on the previously unexamined connection between schools and gentrification. The discovery of this remarkable relationship calls for a deeper analysis. More specifically, future research should investigate whether the relationship is causal. Our hypothesis is that schools stimulate gentrification by attracting younger and more affluent households to the neighborhoods. However, this hypothesis requires a thorough causal analysis. Furthermore, one may also explore whether school characteristics are of any significance in predicting or causing gentrification. In this study, we considered all the schools to be the same irrespective of whether they are public or private, their quality of education, and whether they are primary or high schools. We believe the model can benefit from incorporating such variables in future work.

**Appendix 1**

Socioeconomic feature coefficients from the explained logistic regression model.

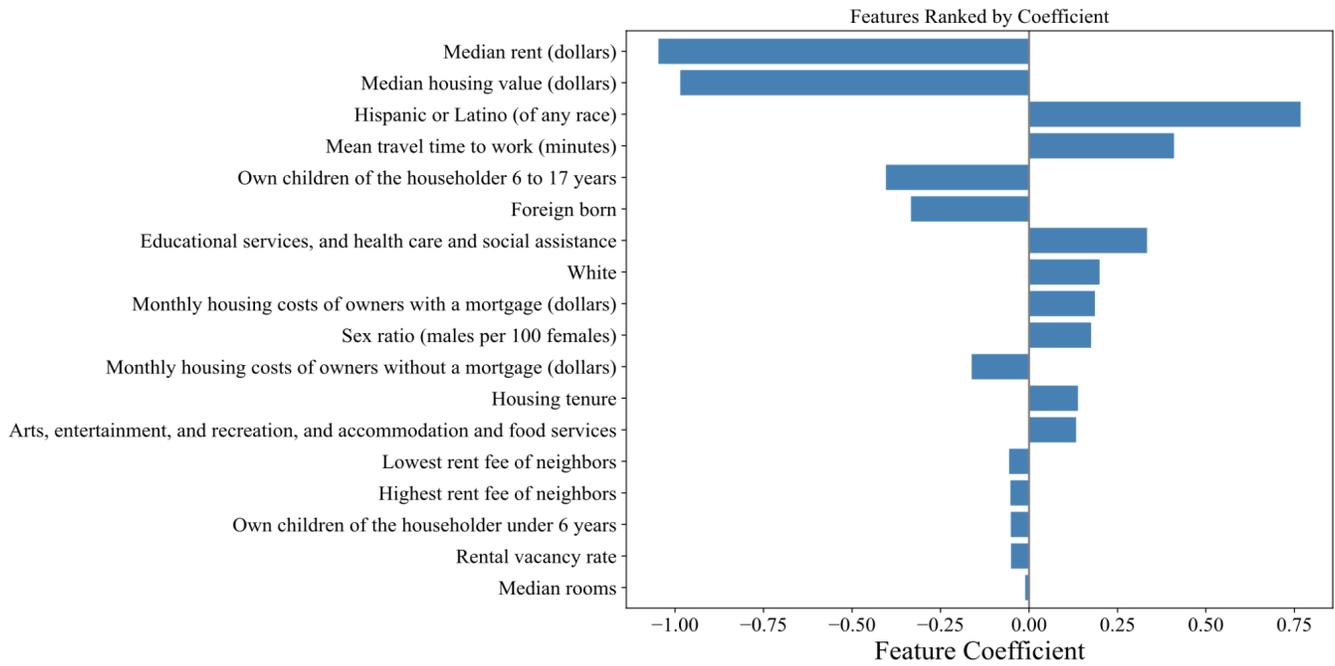